\documentclass[nonacm, sigconf, twocolumn]{acmart}
\renewcommand\footnotetextcopyrightpermission[1]{}
\setcopyright{none}

\makeatletter
\def\@ACM@checkaffil{
    \if@ACM@instpresent\else
    \ClassWarningNoLine{\@classname}{No institution present for an affiliation}%
    \fi
    \if@ACM@citypresent\else
    \ClassWarningNoLine{\@classname}{No city present for an affiliation}%
    \fi
    \if@ACM@countrypresent\else
        \ClassWarningNoLine{\@classname}{No country present for an affiliation}%
    \fi
}
\makeatother

\usepackage{booktabs}
\usepackage{graphicx}
\usepackage{algorithm}
\usepackage[noend]{algpseudocode}
\usepackage{multirow}
\usepackage{graphicx}
\usepackage{amsmath}

\usepackage{url}

\usepackage{breakurl}

\usepackage{float}

\usepackage{xspace}
\usepackage{subfig}
\usepackage{xcolor}
\usepackage{tabularx,colortbl}
\usepackage[inline]{enumitem}

\usepackage{hyperref}
\hypersetup{
  colorlinks=true,      
  linkcolor=blue,       
  citecolor=magenta,    
  filecolor=cyan,       
  urlcolor=red          
}

\usepackage[compact,small]{titlesec}
\usepackage[font={small}, skip=4pt]{caption}

\usepackage{algorithm, algorithmicx,algpseudocode}
\algrenewcommand\algorithmicindent{0.5em}%

\begin{document}
\sloppy
\date{}

\title{Memory Sharing with CXL: Hardware and Software Design Approaches}

\author{Sunita Jain, Nagaradhesh Yeleswarapu, Hasan Al Maruf, Rita Gupta }
\affiliation{\institution{AMD, Inc.}}
\email{{sunita.jain, nagardh, hasan.maruf, rita.gupta}@amd.com}

\begin{abstract}
  Compute Express Link (CXL) is a rapidly emerging coherent interconnect standard that provides opportunities for memory pooling and sharing. Memory sharing is a well established software feature that improves memory utilization by avoiding unnecessary data movement. In this paper, we discuss multiple approaches to enable memory sharing with different generations of CXL protocol (i.e., CXL 2.0 and CXL 3.0) considering the challenges with each of the architectures from the device hardware and software viewpoint. 
\end{abstract}

\maketitle
\thispagestyle{empty}
\def\thefootnote{}\footnotetext{Presented at the 3rd Workshop on Heterogeneous
Composable and Disaggregated Systems (HCDS 2024)}\def\thefootnote{\arabic{footnote}}

\section{Introduction}
In today's server architecture, compute node and memory are tightly coupled and memory technology is completely dictated by the CPUs. Tight coupling between CPU and memory subsystem restricts flexible server system design with different memory hierarchies which leads to stranded compute, network, and/or memory resources~\cite{osr-disaggregation}. This, eventually, increases a datacenter’s Total Cost of Ownership (TCO). With the increasing push towards decoupling CPU and memory resources, a new paradigm shift is taking place for memory access and data movement in next-generation server architectures~\cite{tpp,pond-azure}. CXL~\cite{cxl} is an open, industry-supported interconnect based on the PCI Express (PCIe) interface that enables high-speed, low latency communication between the host processor and end devices (e.g., accelerators, memory expanders, smart I/O devices, etc.). CXL allows cache-line granular access to the connected devices with a byte addressable memory in the same physical address space. From an OS point of view, CXL-Memory appears to the system as a separate NUMA node where memory access latency is slightly higher (50—100 ns) over remote NUMA access latency~\cite{tpp,pond-azure}. This NUMA-like behavior with main memory-like access semantics makes CXL-Memory a good candidate for the slow-tier in datacenter memory hierarchies. 

CXL 1.0/1.1 enables a point-to-point link between CPU and end device. CXL 2.0 allows one-hop switching that enables multiple end-devices to be configured to a single host and supports memory pooling. CXL 3.0 adds multi-hop hierarchical switching – one can have any complex types of networks through cascading and fan-out. This expands the number of connected devices and the complexity of the fabric to include non-tree topologies, like Spine/Leaf, mesh- and ring-based architectures. CXL 3.0 supports PCIe 6.0 (64 GT/s i.e., up to 256 GB/s of throughput for a x16 duplex link) and expands the horizon of very complex and composable rack-scale server design with varied memory technologies. Besides memory pooling, CXL 3.0 enables memory sharing across multiple hosts on multiple end-devices where hardware can manage coherency of the shared memory regions through Back Invalidation (BI). Connected devices (i.e., accelerators, memory expanders, NICs, etc.) can do peer-to-peer communication bypassing the host CPUs. 


\begin{figure}[!t]
    \includegraphics[width=\columnwidth]{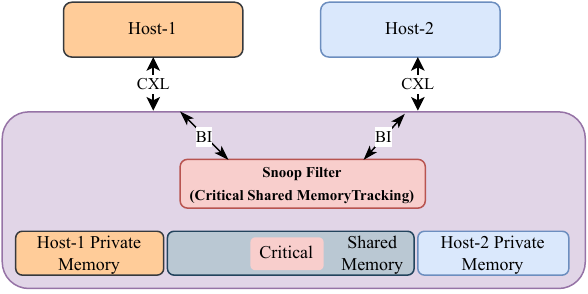}
    \caption{Memory sharing with CXL 3.0 (Type-3 HDM-DB/FAM)}
    \label{figure:memory_sharing}
\end{figure}

From CXL 2.0 onward, devices can be hot-plugged or removed. Device-attached memory is mapped to the system’s coherent address space and accessible to the host using standard writeback semantics. Memory located on a CXL device can either be mapped as Host-managed Device Memory (HDM) or Private Device Memory (PDM). To update the memory address space for connected devices to different host devices, a system reset is needed; traffic towards the device needs to stop to alter device address mapping during this reset period. An alternate solution to avoid this system reset is to map the whole physical address space to each host when a CXL device is added to the system. The VMM or fabric manager in the CXL switch will be responsible for maintaining isolation during address-space management. In case of memory sharing, as CXL 2.0 doesn’t provide hardware support, the software layer needs to ensure all the coherency and consistency. 

CXL 3.0 adds support for BI to maintain hardware coherency for HDM. This allows the device to implement a snoop filter to track HDM lines that are cached in peer caches. As the HDM or shared Fabric Attached Memory (FAM) size is expected to grow, a precise snoop filter that tracks every 64B cache line may turn out to be an impractical solution. On the other hand, using an imprecise snoop filter that tracks at a higher granularity (e.g., 4KB) may issue unnecessary BI snoops and degrade performance. In this regard, one can use a hybrid mechanism – employ a precise snoop filter to track certain region(s) of HDM memory (these shared regions are hardware coherent and can be used for atomics, semaphore, or metadata like queue-pairs, etc.) and the coherency for the rest of the regions can be maintained through the software. This hybrid model allows one to use BI snoop mechanism for a critical portion of memory within a reasonable performance degradation. Figure~\ref{figure:memory_sharing} shows the hybrid mechanism for memory sharing in a CXL 3.0-enabled system.

From the above discussion, CXL 2.0 does not provide hardware support for memory sharing -- memory sharing is only possible through software-based solutions. Although CXL 3.0 has hardware support, for the sake of performance, for a large shared address-space, a significant portion of the memory regions require software support for efficient enablement of memory sharing. In the next sections of this paper, we present different approaches for software only and software-hardware co-design for the enablement of memory sharing over upcoming CXL systems.

\section{Software-Enabled Memory Sharing}


\begin{figure}[!t]
    \includegraphics[scale=0.65]{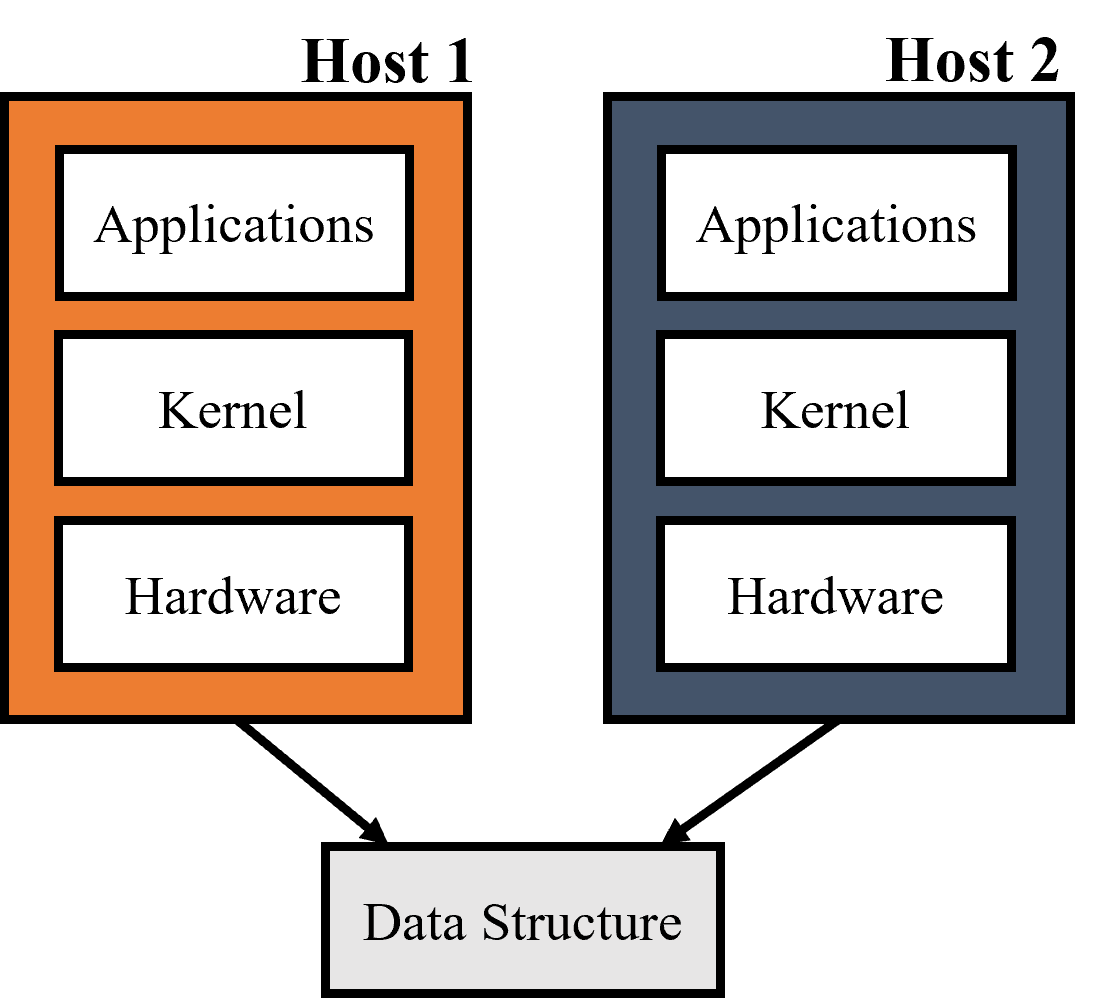}
    \caption{Dual-headed topology for memory sharing}
    \label{figure:dual_head}
\end{figure}

The concept of shared memory is well established in Linux where it is used for Inter-Processor Communication (IPC) and widely deployed in debuggers. It enables efficient data exchange and synchronization between processes without the need for complex data transfer methods such as message passing or file I/O. Shared memory provides a fast and direct means of communication, making it ideal for scenarios where processes need to share large amounts of data or collaborate closely. In Linux, shared memory is implemented through special system calls and functions where processes create shared memory segments, attach them to their virtual address-space with appropriate permissions, read from and write to them, and detach from them when they are no longer needed. 

Linux provides several methods for creating and managing shared memory segments, such as the \texttt{System V} shared memory API, which includes functions like \texttt{shmget}, \texttt{shmat}, \texttt{shift}, and \texttt{shmctl}. Besides, the POSIX shared memory API offers similar functionality through functions such as \texttt{shm\_open}, \texttt{mmap}, and \texttt{munmap}. However, these Linux mechanisms are limited to multiple processes within a single host and cannot be directly used for memory sharing on a multi-headed CXL-system as presented on Figure~\ref{figure:dual_head}.  

The prototype proposed in Figure 2 uses multi-headed memory expansion CXL devices (Type-3 device) connected to multiple hosts. In a software-managed shared memory management solution, memory sharing across multiple hosts can be done either through a custom framework (custom applications/drivers) or an extension of sophisticated libraries like \texttt{OpenSHMEM}~\cite{openshmem}. 

\paragraph{\bf{Custom Framework.}} In such a solution, applications or kernel device drivers need to be re-designed considering the memory sharing feature for underlying CXL devices. Participating hosts (Host 1 and 2 in Figure~\ref{figure:dual_head}) will run their own instances of application/kernel driver. These instances will communicate with each other to maintain the sharing semantics. For example, during the initial memory allocation, instances running on both hosts will agree upon the address range of fixed memory locations in shared region to allocate common data structures. Later, they can access or modify the agreed upon memory addresses to maintain the sharing semantics (e.g., semaphore, synchronized remote memory access, etc.). All the instances will use reserved memory addresses in shared region to exchange information between them.  


\begin{figure}[!t]
    \includegraphics[width=\columnwidth]{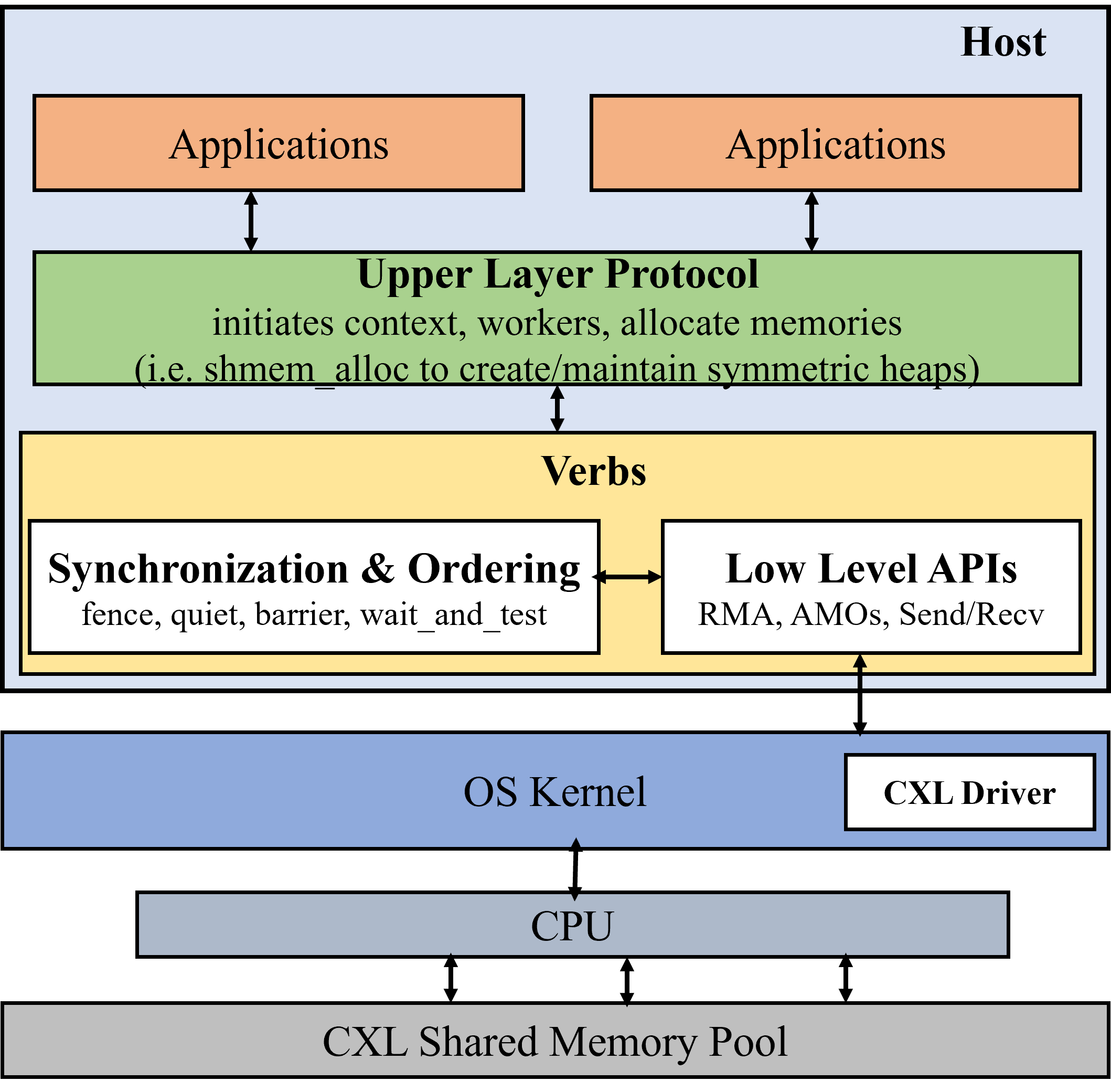}
    \caption{Software stack for custom memory sharing framework}
  \label{figure:software_stack}
\end{figure}

Figure~\ref{figure:software_stack} presents the software stack for designing such a custom framework. The CXL Driver at the OS layer provides support for enabling CXL devices in the system including bringing up the device, populating the memory address-space for the hardware, low-level CXL protocol implementation, support for multi-headed devices, etc. The Verb layer (custom modified for CXL support) in the user-space provides an abstraction of the underlying CXL protocols. It exposes a low-level API to the application developers to implement higher-level sharing-semantics, such as, atomic operations, synchronizations, message orderings and scheduling, etc. The Upper Layer Protocol provides the applications a runtime for memory sharing. 

It creates contexts for each application that provides a conceptual separation between multiple processes along with a support for isolation of communication resources to avoid any interference. Within a context, there can be multiple workers that are responsible for the shared-memory access request generations, message queueing, resource monitoring, memory address-space management, etc. The workers can communicate between each other locally and remotely to track the progress of any shared memory operation. At the initialization of a context, the shared memory regions across different hosts are defined and the workers maintain the global/local memory heaps.

\begin{figure*}[!t]
    \includegraphics[width=0.8\textwidth]{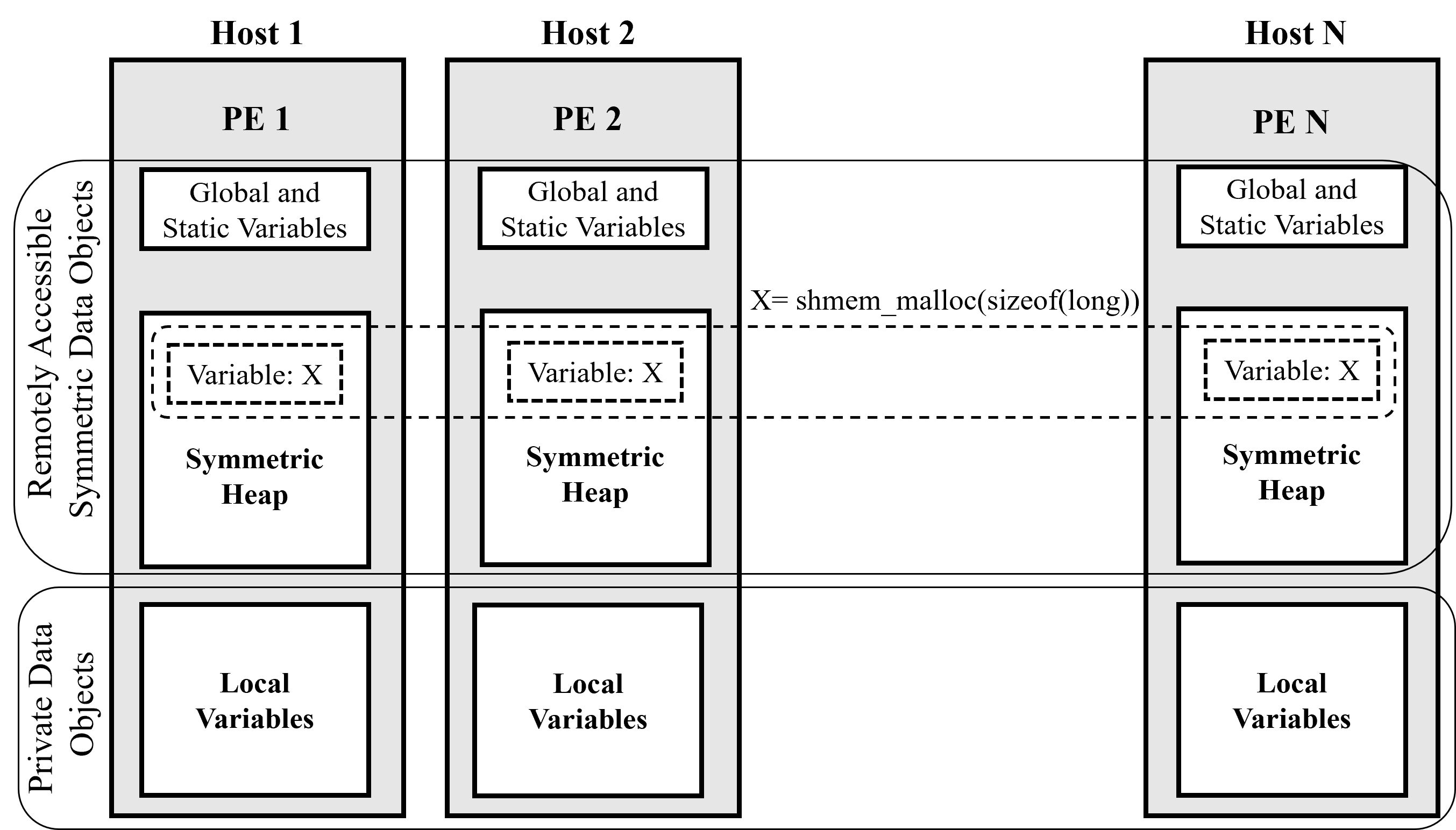}
    \caption{Objects in \texttt{OpenSHMEM} framework}
  \label{figure:openshmem}
\end{figure*}

\paragraph{\bf{OpenSHMEM-based Implementation.}} \texttt{OpenSHMEM} is an open-source and widely used Partitioned Global Address Space (PGAS) library interface specification that aims to provide a standard API for applications to communicate across shared memory regions. \texttt{OpenSHMEM} implementation provides the interfaces for Atomics, Remote Memory Access (RMA), Collectives, Symmetric Memory, and Utils components that are responsible for all data transfer over the shared memory network. Objects within \texttt{OpenSHMEM} implementation can be either private to each Processing Engine (PE) or globally sharable across all the PEs running in remote hosts (Figure~\ref{figure:openshmem}).  

\begin{figure*}[!t]
    \centering
    \subfloat[][{Dual-headed CXL Type-3 device}]{
	\label{fig:multi-headed}
		\includegraphics[width=\columnwidth]{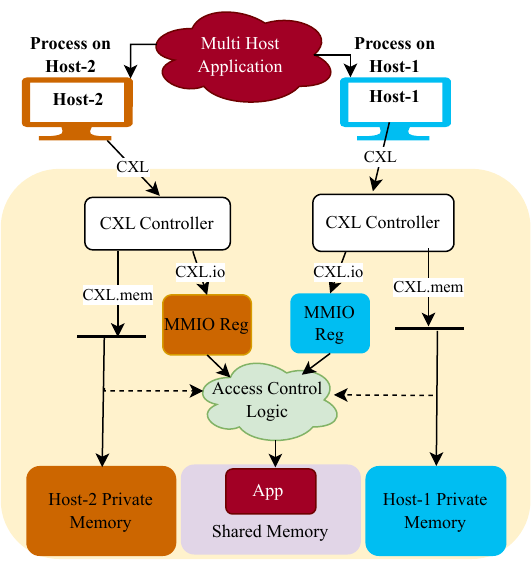}
    }
    \subfloat[][{End-to-end control flow}]{
	\label{fig:control_flow}
		\includegraphics[width=\columnwidth]{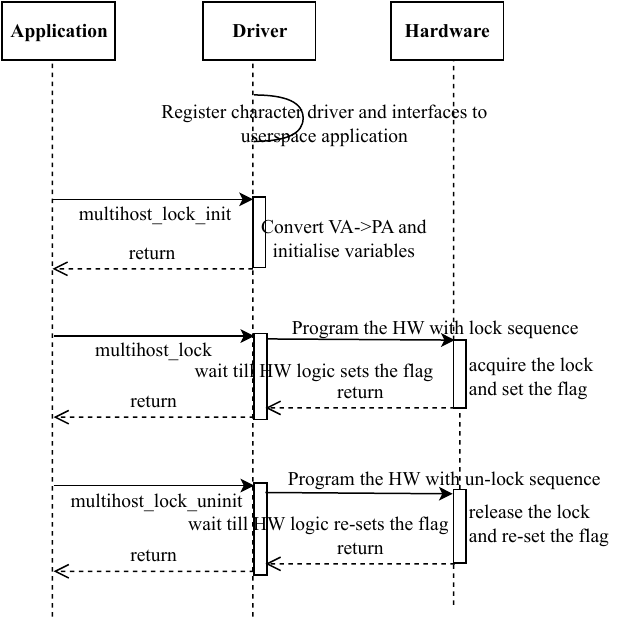}
    }
    \caption{Memory sharing on a dual-headed system}
\end{figure*}
Remotely accessible objects are symmetric across all hosts – all PEs have corresponding objects with the same name, type, and size. It provides pointer routines to query a local address to a remotely accessible data object. \texttt{OpenSHMEM} implementation exposes fencing and barrier functionalities that can be used to implement different programming functionalities for atomic operations over the shared memory region. \texttt{OpenSHMEM}-based shared memory implementation over CXL can be a generic and widely adoptable solution as a lot of details with respect to shared region have already been handled by the framework. 

\section{Hardware-Enabled Memory Sharing}

In this section, we will discuss a hardware approach for shared memory management. Our proposed hardware-based prototype design considers dual-headed CXL Type-3 devices based on a test chip. A dual-headed device has two CXL ports that can connect to two different hosts through appropriate connectors and cabling. Hardware atomics (Test \& Set or Compare \& Swap opcodes) are implemented for shared memory access control via CXL.io Memory Mapped IO (MMIO) region. By design, for a shared memory region, a requester is not expected to issue writes till it grabs the permission for write access. Read is allowed without atomics, i.e., a single writer and multiple reader profile is supported.

CXL controllers and shared memory management mechanisms can be implemented through an FPGA (Figure~\ref{fig:multi-headed}). Each CXL port has its own set of atomics registers implemented in the MMIO space on CXL.io. The Access Control Logic layer provides additional hardware logic that is responsible for granting locks, semaphore, and other access control functionalities.

The hardware design also provides address remap support per controller to map the host physical address to device HDM address. 
Same address remap scheme applies to addresses in shared memory region for hardware atomics. 
In such a hardware-assisted shared memory implementation, the driver is responsible for the address management and providing abstraction over the hardware logic. Figure~\ref{fig:control_flow} shows the control flow between the application, device driver, and the FPGA during a shared memory operation. When an application appears to the system, the shared memory driver registers it and exposes access to the interfaces for hardware functionalities. It then grants access to the memory region that can be shareable to the application and stores all the metadata including virtual to physical address translation, access granularity, permission type, etc. 

When an application appears to the system, the shared memory driver registers it and exposes access to the interfaces for hardware functionalities. It then grants access to the memory region that can be shareable to the application and stores all the metadata including virtual to physical address translation, access granularity, permission type, etc. When the user space application requests for a write access permission, the driver sends a message to the hardware to initiate hardware atomics. Once the access permission is acquired and the corresponding flag is set, the driver returns to the application. All the write access requests are ordered and served based on the arrival sequence. The write access release mechanism follows a similar routine – hardware atomics is issued and the flag is reset, the application is notified, and the next write access request can be executed. 

\section{Discussion}
\paragraph{\bf{Sharing Granularity.}} Memory sharing improves application-level performance by reducing unnecessary data movement and improves memory utilization. Sharing a large memory address space, however, may result in significant overhead and complexity in the system. For example, in a terabyte or even larger system, sharing the whole address space across multiple hosts with high access frequency may cause high overhead due to consistency management. Continuously invalidating and updating the memory regions across all the hosts can be extremely expensive operations. Besides, cache line or page granular lock management can be impractical because of the memory overhead due to metadata management. This overhead can be reduced if we consider larger memory regions (e.g., 1GB region). However, if multiple hosts have frequent write request rate, such large granular regions may add high wait time for other hosts to grab the lock. CXL-enabled multi-headed shared memory systems will be ideal for applications that have high read and low write demands.    

\paragraph{\bf{Security.}} Furthermore, sharing memory across multiple devices increases the security threat in the presence of any malicious application run on the same hardware space. When an application disappears from the system, assigning a fraction of its previously shared memory space to a newly arrived application may leak sensitive information if proper encoding, isolation, and clean up mechanism is not exercised. Hardware assists can be deployed to zeroize the memory region before assigning the same to a new application. 

\section{Conclusion}
In this paper, we present different approaches for enabling CXL-based memory sharing with different generations of CXL protocol and discuss their trade-off. With CXL 3.0 features the entire rack can share memory. This becomes very powerful with Global Integrated Memory (GIM). In addition, near data processing with memory sharing will open up new use-cases for CXL at rack-scale. 

\label{lastpage}
\bibliographystyle{abbrv}
\bibliography{reference}

\end{document}